\documentclass[conference]{IEEEtran}
\usepackage[cmex10]{amsmath}
\usepackage{siunitx}
\usepackage[font=footnotesize]{caption}
\usepackage{psfrag}
\usepackage[utf8]{inputenc}
\usepackage[T1]{fontenc}
\usepackage{amsmath,amsfonts,amsbsy,amssymb}
\usepackage{mathabx}
\usepackage{mathrsfs}
\usepackage[nolist]{acronym}
\usepackage{tabularx}
\usepackage{amssymb}
\usepackage{pdfpages}
\usepackage{amsmath}
\usepackage{graphicx}
\usepackage{cite}
\usepackage{multirow}
\usepackage{wasysym}
\usepackage{multirow}
\usepackage{float}
\usepackage{xcolor}
\usepackage{subfig}

\usepackage[ruled, lined, longend, linesnumbered]{algorithm2e}

\usepackage{xcolor}
\usepackage{url}
\usepackage{multicol}
\usepackage{verbatim}
\usepackage{array}

\title{A Heuristics-based Home Energy Management System for Demand Response }

\author{Hafiz Majid Hussain and Pedro H. J. Nardelli\\
Department of Electrical Engineering\\School of Energy Systems, LUT University, Finland
}

\begin{document}

\maketitle
\thispagestyle{empty}
\pagestyle{empty}


\begin{abstract}
The so-called Internet of Things (IoT) and advanced communication technologies have already demonstrated a great potential to manage residential energy resources via demand-side management.
%
%
%
This work presents a home energy management system in that focused on the energy reallocation problem where consumers shall shift their energy consumption patterns away from peak periods and/or high electricity  prices. 
Our solution differentiates residential loads into two categories: (i) fixed power appliances and (ii) flexible ones.
Therefrom, we formulate our problem as constraint optimization problem, which is non-linear and cannot be mathematically solved in closed-form.
We then employ and compare two well-known heuristics, the genetic algorithm (GA) and the harmony search algorithm (HSA), to minimize electricity expense and peak to average ratio.
These two approaches are compared to the case where no reallocation happens.
Our numerical results show that both methods; GA and HSA can effectively reduce the  electricity cost by 0.9\%, 3.98 \%,  and  PAR  by  15\%,  5.8\%, respectively. 
%
\end{abstract}

\section{Introduction}
%
%
Smart grid technologies, smart meters and demand response have enabled consumers to know their demand profile in more details, while helping the system operator to improve the efficiency and reliability of the power system \cite{01}.
In particular, demand-side management (DSM) has great impact on grid operation by, for example, facilitating the incorporation of renewable resources, and by allowing the consumers to actively participate in electricity dispatch. 
As part of this broader concept, demand-response (DR) is defined as \cite{02}: ``Changes in electric usage by end-use customers from their normal consumption patterns in response to changes in the price of electricity over time, or to incentive payments designed to induce lower electricity use at times of high wholesale market prices or when system reliability is jeopardized''. 

In this context, DR can categorized into two aspects, viz. incentive and price-based programs\cite{03}.
Incentive based programs involve customers' participation to reallocate their energy consumption in off peak hours in response to which a reward (bill credit payment) is given to them for their participation in the program.
Incentive programs are direct load control (DLC), curtailable load, demand bidding \& buy back, emergency \& demand. 
On the other hand, price-based programs involve various pricing signals at different times to reduce energy consumption by providing monetary benefits to the consumers.
It includes time of use, real time pricing, inclined block rate, critical peak pricing and day ahead pricing \cite{04}.  In a recent research, price-based DR has been studied widely in residential sector, particularly, in home energy management system (HEMS).
For instance \cite{04,05,06,07,08}, various HEMS models in the context of DR have investigated to achieve optimal energy consumption of household appliances using optimization model, aiming to reduce electricity cost, balance energy demand, and improve energy efficiency. 
%

In general, HEMS plays a significant role in energy management of residential sector and allows exchange of energy consumption information with the utility to improve energy profile as well as the reliability of power grid. 
The work in \cite{04} comprehensively described HEMS architecture, DR programs, smart grid technologies, communication protocols, and various decision making algorithms like
artificial intelligence (AI) and heuristic scheduling algorithms. 
These algorithms are considered as an essential part towards the energy optimization and load shifting operations in HEMS.
Fan-Lin and Xiao-Jun in \cite{05}, designed a residential energy usage framework using genetic algorithm (GA) which attempts to maximize re-trailer's profit. The home appliances are classified in two groups (shift able and curtail able) and hourly energy usage is predicted in accordance with the electricity price and temperature signal.
In another work \cite{06}, a multi-objective problem is applied to control energy consumption of household micro-grid and hybrid differential evolution is used to solve the scheduling problem.
In a similar context with a recent work \cite{07} authors have been explored the HEMS based on hybrid optimization technique to manage energy consumption of smart appliances in \textit{24} hours time slots depending upon pricing tariffs and coordination among appliances.
In \cite{08}, authors extended their previous work and incorporated various time slots, peak to average ratio (PAR), and multiple homes scenarios with much improved hybrid technique bacterial flower pollination algorithm (BFPA). 

Kai Ma et al. further developed an optimization problem in \cite{09} and investigated the trade-off between electricity cost and discomfort cost.
In \cite{10}, a generalize HEMS discussed based on GA to schedule energy consumption and minimize operational cost of electricity considering user satisfaction constraints. 
Similarly, reference \cite{11} adopted GA based on DSM (GA-DSM) strategy to distribute the power in industrial area effectively. In an other contribution  \cite{12}, authors interested to analyze scheduling mechanism in domestic sector using  binary particle swarm optimization (BPSO) to optimize energy consumption of household in pre define time intervals. 
Different from GA and BPSO, authors in \cite{13} have been proposed an improved algorithm binary backtracking search algorithm (BBSA) to balance energy usage and effectively control cost.
The simulation results of BBSA and BPSO are compared which shows effectiveness of BBSA.
In the same fashion authors in \cite{14,15} introduced practical pricing and green energy scheduling plan with an aim  of minimizing overall electricity cost while applying different approaches such as, non linear programming and game theory algorithm, respectively.
In the same sense, but with different approach, we develop here an efficient DR strategy to lessen the cost and the peak-to-average ratio of energy usage, which is expect to contribute towards the green house emissions and fuel wastage. 
%

Our goal here is to explore the energy consumption behaviour for residential consumer in order to shift some specific loads trying to shape the load curve, accordingly. 
This paper proposes an optimization model for scheduling energy consumption of various kind of appliances, which are classified into two groups based on their features and parameters, namely fixed power and flexible power appliances.
We compare the performance of three different approaches: optimization via GA, optimization via harmony search algorithm (HSA), and no optimization (i.e., no load shift is performed).
Our results compare the cost and the peak-to-average ratio in the three scenarios, showing that designed algorithms have the best performance. Our contributions are summarized as:
\begin{itemize}
\item We develop  DR strategy to  address the  peak load shaving problem and flexibly control the household appliances specifically, at times when prices of electricity are high.
\item To address the problem, we develop system model considering the household appliances and classifying them into two types; fixed power and flexible power appliances. The energy consumption of the appliances are managed and controlled considering the time of use pricing model.
\item In order to solve the problem, we establish optimization model along with two well known heuristics; GA and HSA. 
\item We analyze three different scenario: Without HEMS, With HEMS-GA, and With HEMS-HSA.  For example, with  HSA-HEMS the cost and PAR are reduced to 3.98\% and 5.8 \%, respectively.
\item We demonstrate the proposed  solution is scalable for various scenarios by testing the designed algorithm with multiple users case i.e., 10 users and 50 users considering  different time resolutions (60 minutes and 30 minutes).
\end{itemize}

 The rest of the paper is organizes as follows. 
 Section II states the system model used here, including the problem formulation, its input parameters and the optimization methods used.
 Section III presents the numerical results and the performance evaluation of the three different scenarios.
 Section IV conclude this paper and propose some potential future work.

\section{System Modeling}

In this work, we investigate the energy reallocation problem of peak hours based on DR strategy.
We consider home energy management environment where each home is equipped with HEMS with the function of optimizing energy consumption of household appliances based on different input parameters as electricity price and type of load.
%
%
The two way of communication between HEMS and utility enabled the consumers to alter the energy usage based on electricity price signal.
The electricity price depends on the demand of energy, higher the demand the higher will be the electricity cost and vice versa. 
The demand information of energy is transmitted by smart meters to utility via IoT network. 
The peak energy demand (of home appliances) can be controlled appropriately by addressing the PAR. 
So that, we assorted home appliances into two types based on their features and priorities, namely, fixed and flexible loads\cite{19,16}. 
%



In order to support the communication infrastructure,   advance metering infrastructure (AMI) is an essential element in smart grid. 
AMI combines multi-way communication, data management system and particularly, smart metering system. 
This enables smart meters (SM) to measure and collect the information of energy consumption in an accurate and precise way.
Moreover, this information is also exchanged between HEMS and utility industry simultaneously in a real time scenario. The communication between HEMS and utility industry also enables the user to take part in DR strategies and manage the energy demand effectively.
On the other side, users in home can monitor the information such as available energy, energy consumption, price of energy in the next hour, etc., using various interfaces e.g., smart phones, computers etc., and adjust energy consumption based on a DR strategy. 
%

\subsection{Appliances classification}
Appliances are classified here into two types, namely fixed power and flexible power appliances, as discussed next.

\subsubsection{Fixed power appliances} $(\mathcal{A}_{\scalebox{.85}{$\scriptstyle\mathcal{S}$}}^{\scalebox{.85}{$\scriptstyle {fixed}$}})$
fixed power appliances are ceiling fan, lamp,  or TV, these have fixed power consumption profile and operational time and due to continuous power supply  HEMS will not  schedule fixed power appliances i.e.,
$\mathcal{A}_{\scalebox{.85}{$\scriptstyle\mathcal{S},{h}$}}^{\scalebox{.85}{$\scriptstyle {fixed}$}}=E_{fixed}$

\subsubsection{Flexible power appliances} $(\mathcal{A}_{\scalebox{.85}{$\scriptstyle \mathcal{S}$}}^{\scalebox{.85}{$\scriptstyle{F}$}})$ Flexible power appliances can be controlled and their energy consumption profile are scheduled by HEMS.
Their operation is attributed as incentive-based $(\mathcal{A}_{\scalebox{.85}{$\scriptstyle \mathcal{S,I}$}}^{\scalebox{.85}{$\scriptstyle {F}$}}$) and price-based (\(\mathcal{A}_{\scalebox{.85}{$\scriptstyle \mathcal{S,P}$}}^{\scalebox{.85}{$\scriptstyle {F}$}}\)).
The energy usage of (\(\mathcal{A}_{\scalebox{.75}{$\scriptstyle \mathcal{S,I}$}}^{\scalebox{.75}{$\scriptstyle \mathcal{F}$}}\)) is curtailed considering DR strategy.
Various pricing signals can be adopted to reallocate the load demand from peak to off-peak hours to achieve cost reduction. The price based flexible appliances are of two types (i) non-interruptible and (ii) interruptible.
The operation time interval of non-interruptible appliance must not be halted during their operating time by HEMS such as, washing machine and iron.
Interruptible appliances can be interrupted in any time period like, during the peak demand or high cost of electricity generation e.g., air condition and water heater. The energy usage of interruptible appliances are presented below: $E_{{flex}_{\scalebox{.85}{$\scriptstyle {min}$}}} \leq E_{flex} \leq E_{{flex}_{\scalebox{.85}{$\scriptstyle {max}$}}}$. The power rating (PR) and operational time interval (OTI) are shown in the Table \ref{table_app}

\subsection{Electricity  pricing model}
\begin{figure}[t]
     \centering
     \includegraphics[width=1\columnwidth]{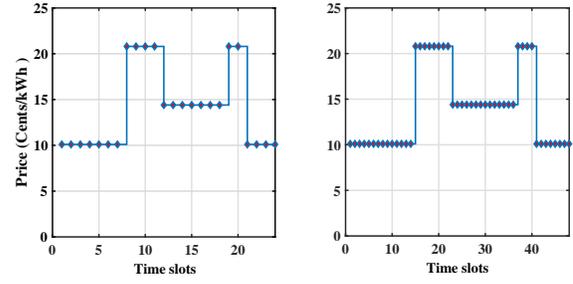}
     \caption{TOU electricity price tariff, left to right side: price for one day with time slots 60 and 30 minutes.}
     \label{fig_tou}
 \end{figure}

Pricing tariff refers to various pricing scheme for designated time frame. DR based pricing tariff plays important role to allow active participation of consumer in residential sector. 
Among various pricing tariffs discussed in the literature \cite{03,04,16}, we opted time of use pricing model in our simulation results. It is briefly discussed below:

\textit{Time of  use}: Time of use (TOU) pricing scheme reflects price of electricity in different time of interval including, off peak, mid peak, and peak hours. 
TOU tariff imparts the average electricity cost of power generation during different time periods and allows the consumers to manage their energy usage voluntarily instead of being forced by utility.
In the same way, consumers have the flexibility either to use the electricity in peak time interval (which yields higher cost) or off peak (lower cost due to less stress on generation resources). Typically, TOU is spreading widely and used in many countries for residential sector consumers.
For instance, TOU tariff is implemented in USA, Canada, and Ireland, and customers pay their bill according to fixed prices in different time periods i.e., during  off , mid, and peak hours\cite{03,20,21}. 
We have used TOU taken from \cite{22}.
An example of TOU is given in Fig. \ref{fig_tou}.

The total cost of electricity can be expressed using ToU pricing $\gamma$, and states of the household appliances $\pi$ as:
\begin{equation}
\mathcal{C}_T(t)= \sum_{t=1}^\mathcal{T}\mathcal{E}(t) \times \pi(t) \times \gamma (t).
\end{equation}

\subsection{Cost function and energy demand}
Let $\mathcal{A_S}$ represents set of appliances and $\mathcal{P}_{\scalebox{.85}{$\scriptstyle{fixed}$}}(t)$, $\mathcal{P}_{\scalebox{.85}{$\scriptstyle{F}$}}(t)$ denote the energy consumption of fixed and flexible power appliance in time $(t)$. 
The total energy consumption $(\mathcal{P}_{\scalebox{.85}{$\scriptstyle{T}$}}(t))$ in each time period $t$ $\epsilon$ $\mathcal{T_S}$=\{1,2,....$\mathcal{T}$\}, then considering this definition total energy usage during $t$ $\epsilon \mathcal{T}$ can be calculated as $\mathcal{E}(t)=\sum_{\substack{\scalebox{0.5}{$\mathcal{A}=1$}}}^{\substack{\scalebox{0.5}{$\mathcal{A_S}$}}}$  $(\mathcal{P}_{\scalebox{.85}{$\scriptstyle{T}$}}(t))$. The overall cost is expressed mathematically as:
\begin{equation}
E_T=\sum_{{t}=1}^{\mathcal{T}}\sum_{\mathcal{A}=1}^{\mathcal{A_S}}\bigg(\mathcal{E}(t)\times \pi(t)\times\gamma(t)
\bigg).
\end{equation}        

In above equation, the first term on the right side computes the cost of electricity in each time slot $t$; the second term computes amount of energy used in t-th hour of the day; $\pi$ is the decision variable that represents ON and OFF states of the appliances. As we are interested in reducing electricity cost, nevertheless the reallocation of energy into off peak hours is also an imperative step to improve the functionality of the grid. Therefore, PAR is computed as:
\begin{equation}
PAR=
 {\frac{G_{peak}}{G_{avg}}}={\frac{\mathcal{T} \underset {t\epsilon\mathcal{T}}{max} \hspace{0.5 em}{\mathcal{E}}(t)}{\sum_{t=1}^{\mathcal{T}} \mathcal{E}(t)}},
\end{equation} 
where $G_{peak}$ and $G_{avg}$ indicate the maximum and average aggregated load in any time slot ($t$).

\subsection{Objective function}
In general, the focus of this work is to jointly minimize the cost of energy and PAR. To accomplish the objective, HEMS is considered to schedule the energy usage of $\mathcal{A_S}$ using optimization problem. 
\begin{center}
Objective 
$\begin{cases}
       \text{$E_T$}\\
      \text{$PAR$}\\
      \text{$\mathcal{E}$(t) (scheduling)}
\end{cases}$
\end{center}
The $E_T$ represents the total energy usage cost; PAR is the fraction of  maximum aggregated energy consumed and mean value of the total energy. The constraints related to objective are as follow:

\begin{equation}
\sum_{{t}=1}^{\mathcal{T}}{\Psi_{\scalebox{.85}{$\scriptstyle{With-HEMS}$}}} \leq \sum_{{t}=1}^{\mathcal{T}} {\Psi_{\scalebox{.85}{$\scriptstyle{Without-HEMS}$}}}
\end{equation}

\begin{equation}
\vspace{1ex}
\gamma_{i,t}=\
\begin{cases}
       \text{1},  \hspace{0.5 em}{\forall  \quad t \in OTI}\\
      \text{0},\hspace{0.5 em} otherwise \\
\end{cases}
\end{equation}
\begin{equation}
\sum_{{t}=1}^{\mathcal{T}}\Upsilon_{{\mathcal{A}_S},{t}}=\Upsilon_{OTI}\hspace{0.5 em} \forall t \in \mathcal{T}
\end{equation}
\begin{equation}
PAR_{\scalebox{.85}{$\scriptstyle{SCH}$}} \leq PAR_{\scalebox{.85}{$\scriptstyle{UNSCH}$}}
\end{equation}
\begin{equation}
0\leq\mathcal{E}(t)\leq\mathcal{G_{T}}
\end{equation}

The equations (4) to (8) represents the constraints of the designed model. 
The constraint (4) illustrates the total cost of the energy  {"With-HEMS"} must be less or equal to {"Without HEMS"}. Constraint (5) shows appliances states, {"1"} indicates ON and {"0"} OFF state. The constraint (6) means the OTI of each appliance should be completed before and after scheduling. The constraint (7) reflects that PAR should be remained less or equal to case (Without-HEMS). The last constraint describes that energy consumption of household should not exceed the total available energy. 
 
\subsection{Optimization techniques}
\subsubsection{Genetic Algorithm (GA)}
is meta heuristic algorithm and inspired by the theory of natural evolution.
 GA is one of the most applied algorithm in various field of computer science and engineering due to the fast computational time and easy  implementation of many complex problems. Among them GA is one of the most applied algorithm in various field of computer science and engineering \cite{23}.
 GA is influenced by biological evolution process which is based on genetic inheritance and natural selection.
 GA is population based heuristic algorithm and starts with the initialization of population then each candidate in the population (known as genes) is evaluated using objective function.
 To select better candidate for the next iteration, we introduce tournament selection. The role of selection is to select best individuals (parent) for recombination and replacement process. Usually, recombination (crossover) and replacement (mutation) are the main driving agents to modify the population and provide diverse search space.
 In our designed model, we implemented GA that is associated with binary representation where {“0"} indicates the OFF and {“1"} shows ON state. Then,  each candidate in the population is tested by objective function. Two point crossover and uniform mutation are introduced to achieve better results. 
After crossover and mutation the new set of candidates again evaluated and compared with previous candidates.
The stopping criteria is maximum number of population size, and the allocation is the best candidate that satisfies the objective function.

\subsubsection{Harmony Search Algorithm (HSA)}
is a popular meta heuristic algorithm inspired from musical improvisation process \cite{24}. It is developed with an aim to search best state of harmony. This (best) harmony in the music is similar to optimization process to find global optimal solutions for a given objective function.
 HSA is an idealising mapping from the qualitative improvisation into quantitative formulation, and hence transforming musical harmony into optimization process.  
The HSA steps are given in the following.
 
\textbf{ Step 1:} In the beginning, HSA parameters are initialized such as, size of harmony memory (HMS), harmony memory consideration rate (HMCR), bandwidth distance (BW), pitch adjustment rate (Par), harmony memory (HM), and total improvisations (NI).
 
\textbf{ Step 2: }In the second step, initial random population is generated using Eq (4). This uniformly random distributed population is stored and analyzed in HM then evaluated using objective function. 
 \begin{equation}
  A_{i,j}^0=   X_j^{min}+B_j(X_j^{max}-X_j^{min}),
 \end{equation}
 where $X_j^{max}$ and ${X_j^{min}}$ are the upper and lower limits and j=1,2,3,..HM. 
 
\textbf{Step 3:} In this step, a set of new vectors known as harmony vectors are generated based on the criteria, HMCR, (1-HMCR), and Par. The stored values in HM are then selected with HMCR and Par probability or it can be opted randomly from HM with the probability of (1-HMCR). In the designed model, it is important to select the best set of candidates from HM in order to effectively minimize objective function. The above discussion can be mathematically expressed as \cite{25}:
 \begin{equation}
 A_{new}=
 \begin{cases}
   A\epsilon\{{a_{1i},a_{2i},a_{3i}\ldots a_{HM}}\},\hspace{0.1cm}With\quad P(HMCR) \\
   A\epsilon\{{a_{1},a_{2},a_{3}\ldots A_{N}}\},    \hspace{0.3cm} With\quad P(1-HMCR)
\end{cases}
\end{equation}

\begin{equation}
A_{new}=
\begin{cases}
  YES, & With \quad P(Par)  \\
  NO, & With \quad P(1-Par).
\end{cases}
\end{equation}

In each iteration this process searches  new best harmony (solution) and replaces the worst individual in HM. 
The process is terminated when stopping criteria (total number of improvisation) is met.

\section{Simulation and experiment results}

\begin{table}[t]
\vspace{1ex}
\centering
\caption{Appliances' characterization}
\label{table_app}
\begin{tabular}{l|l|c|c}

\textbf{Appliance}&\textbf{Class} & \textbf{\small PR (kWh)}& \textbf{OTI (hours)} \\
 \hline
  \hline
         Ceiling fan     & fixed        & 0.075    &14\\ 
         Lamp            & fixed        &  0.1     & 13 \\ 
          TV             & fixed        &0.48      & 7  \\
         Oven            &fixed         &2.3       &6\\
      Washing machine    &flex        & 0.7      & 8\\ 
     Iron                &flex        & 1.8      & 7  \\
       Air conditioner   &flex        &1.44      & 10 \\ 
      Water heater       &flex        & 4.45      &8  \\

\hline
\end{tabular}
\end{table}

\begin{figure}[t]
    \centering
    \includegraphics[width=1\linewidth]{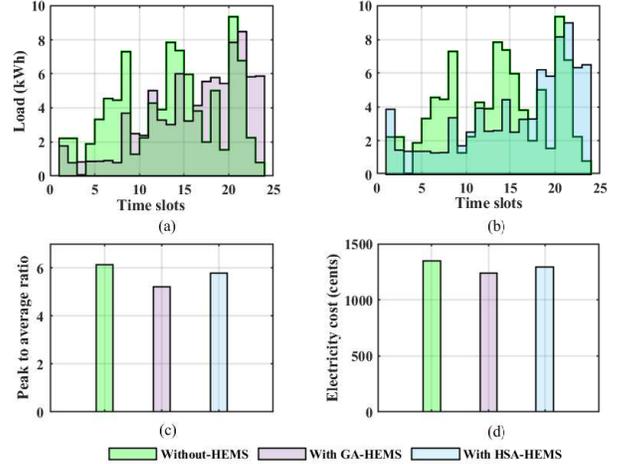}
    \caption{Single user, Legend: (a) and (b) Energy usage profile for 60  minutes time resolution (c) PAR (d) Electricity cost}
    \label{fig:1}
\end{figure}

\begin{figure*}[t]
  \centering
  \subfloat[]{\includegraphics[width=\columnwidth]{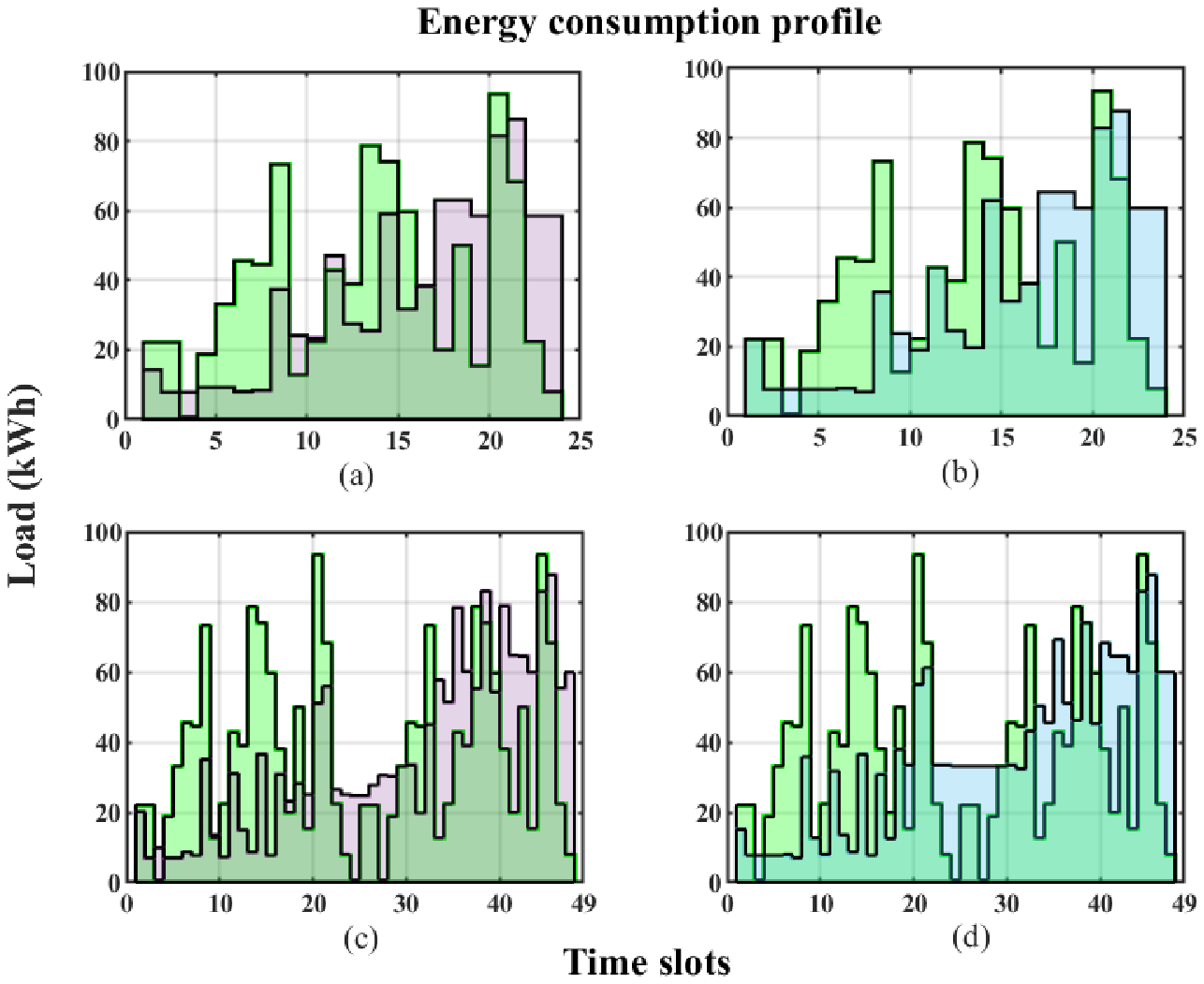}\label{Fig:2(a)}}
  \hfill
  \subfloat[]{\includegraphics[width=\columnwidth]{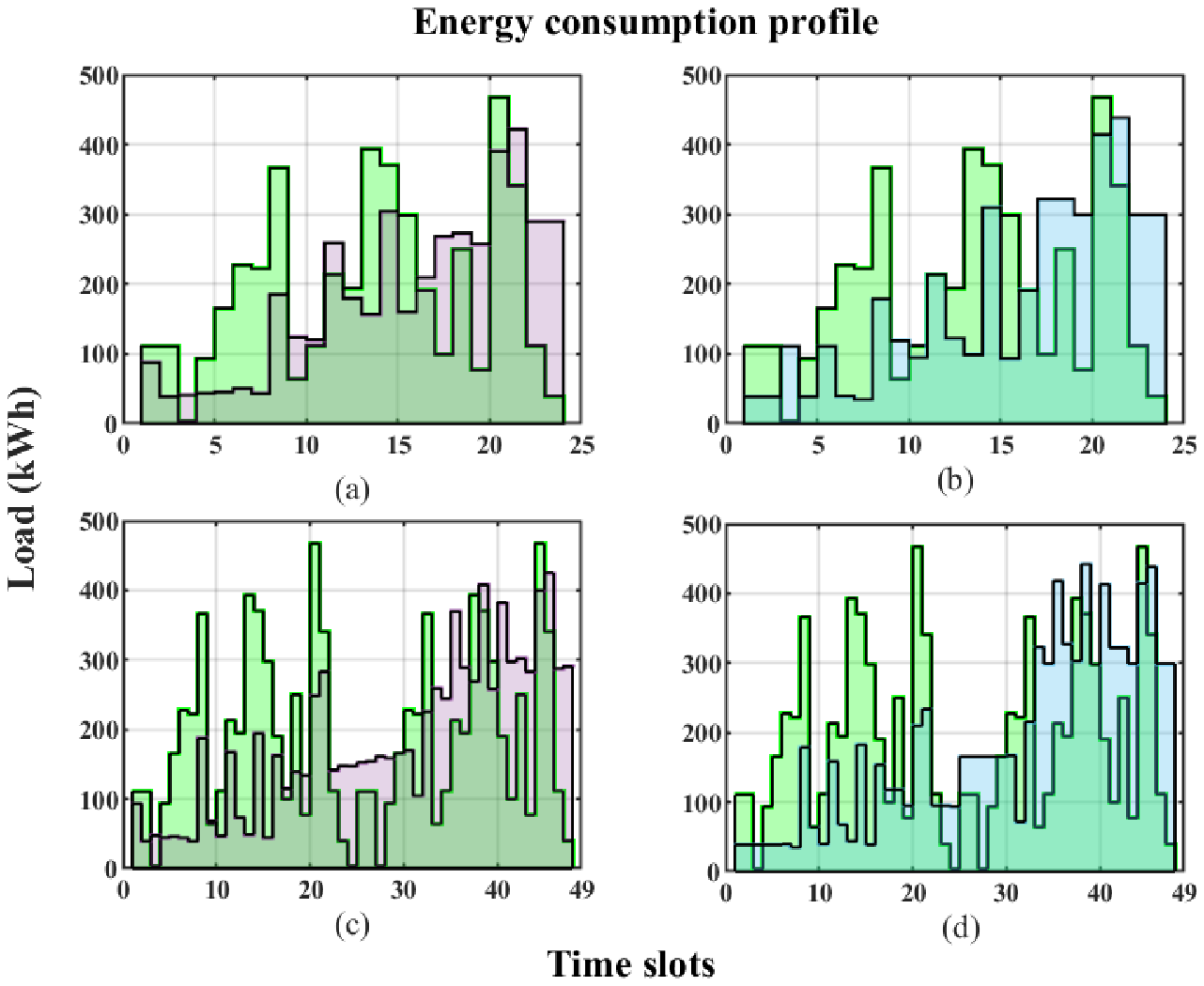}\label{Fig:2(b)}}
  \caption{Energy consumption information (a) 10 users with time slots of 60 minutes and 30 minutes (b) 50 users with time slots of 60 minutes and 30 minutes}
  \vspace{-2ex}
\end{figure*}

\begin{figure}[t]
    \centering
    \includegraphics[width=1\linewidth]{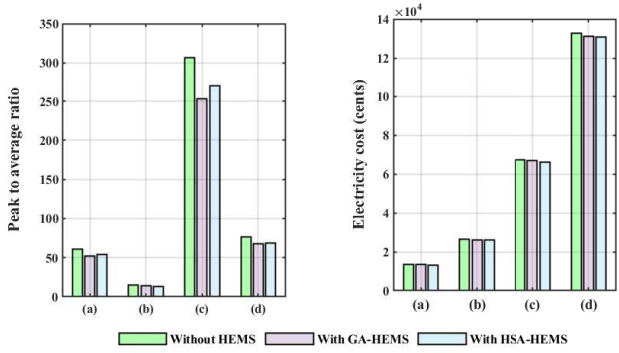}
    \caption{PAR and electricity cost information. Legend: (a)10 users with timeslot 60 minutes (b) 10 users with time slot 30 minutes (c)50 users with time slot 60 minutes (d) 50 users with time slot 30 minutes} 
    \label{fig:3}
    \vspace{-2ex}
\end{figure}

In this section, we  conduct the simulation results based on metrics energy consumption, electricity cost, and PAR.
To present the performance of optimization algorithms, we investigate the experimental results of particularly eight appliances including; four fixed power, two non-interrupt-able, and two interrupt-able flexible power appliances.
The fixed power appliances consume fixed power and cannot be scheduled by HEMS (e.g., fan, lamp, TV, and oven). 
However, non-interrupt-able appliances operate on fixed power and can not stop their operation during scheduling time period (e.g., washing machine and iron) we assume that HEMS will schedule the iron operation after washing machine. 
While, interrupt-able appliances (air condition and water heater) can be controlled and scheduled based on the pricing signal in any time period. %
To present our results, we consider energy consumption of household appliances for one day with time  resolution of one hour ($t$) (starting from 12 am to the next day 12 am) and the TOU pricing tariff for Winter season (November 1, 2018 - April 30, 2019).
%
%
Moreover, we also demonstrate that above scenario can be used for multiple users and various time resolutions such as; 10 users with time resolution 60 and 30 minutes and 50 users with time resolution 60 and 30 minutes.

 As mentioned earlier, energy scheduling is one of the core motivation of this work, therefore, HEMS is designed based on the optimization algorithms; GA and HSA. Fig. ~\ref{fig:1} represents load profile (a and b), PAR (c) , and electricity cost (d)  in one day. It is seen in Fig. ~\ref{fig:1} that each algorithm attempts to schedule energy profile in off peak time (i.e., 21-th hour evening time to 7-the morning time) when the price of energy is low (6.5 cents/kWh). While in peak hours (7  to 11 and 5  to 7 (am)) with price 13.2 cents/kWh, the maximum energy is consumed 8.72 kWh by {"Without HEMS"} whereas, GA-HEMS and HSA-HEMS accounted for 8.67 kWh and 7.14 kWh, respectively. On the other hand, ~Fig. \ref{Fig:2(a)} and Fig. ~\ref{Fig:2(b)} illustrate the  energy consumption of the household appliances for 10 and 50 users with time horizon of 60 and 30 minutes. The maximum energy consumption is 9.35 kWh for single user and 93.5 kWh for 10 users with time span of 60 and 30 minutes while these consumption are flexibly control and shifted to off-peak times in order to reduce the electricity bill. Furthermore, the maximum consumption in case of 50 users (both time interval) are also optimized by GA-HEMS and HSA-HEMS by 9.82 \% and 6.20 \%, respectively. 

\begin{table*}[t]
\centering
\caption{Comparative performance based on numerical results. Costs are in cents (\cent) and Red. means reduction.}
\label{table_consump}
\begin{tabular}{l|c|c|c|c|c|c|c|c|c}

                                                               & \multicolumn{3}{c|}{\textbf{Without HEMS}} & \multicolumn{3}{c|}{\textbf{With GA-HEMS}}                                                            & \multicolumn{3}{c}{\textbf{With HSA-HEMS}}                                                           \\  \hline
\textbf{Designed case}                                                  & \textbf{Max $E_H$}  & \textbf{Cost (\cent)} & \textbf{PAR}    &  \textbf{\% $E_H$ Red.} & \textbf{\% \cent~Red.} &\textbf{ \% PAR Red.} & \textbf{\% $E_H$ Red.} & \textbf{\%  \cent~Red.} & \textbf{\% PAR Red.} \\ \hline
\hline
\textbf{1 user; 60 min}                                                    & 9.35       & 1347.9      & 6.13   & 9.30                            & 0.9                          & 15                          & 3.95                            & 3.98                         & 5.8                         \\ \hline
\textbf{10 users; 60 min}                                                    & 93.50      & 13479       & 61.36  & 7.74                            & 0.71                         & 14.6                        & 6.20                            & 1.37                         & 11.94                       \\ \hline
\textbf{50 users; 60 min}                                                      & 467.50     & 67394       & 306.79 & 9.82                            & 0.98                         & 17.41                       & 6.20                            & 1.83                         & 12.02                       \\ \hline
\begin{tabular}[c]{@{}l@{}}\textbf{10 users; 30 min } \end{tabular} & 93.50      & 26569       & 15.34  & 6.20                            & 1.41                         & 6.51                        & 6.20                            & 1.67                         & 12.05                       \\ \hline
\begin{tabular}[c]{@{}l@{}}\textbf{50 users; 30 min}\end{tabular} & 467.50     & 13285       & 76.70  & 9.00                            & 1.36                         & 11.86                       & 5.40                            & 1.64                         & 10.52                       \\ \hline
\end{tabular}
\label{tab-anali}
\end{table*}

The effectiveness of designed HSA-HEMS and GA-HEMS is also evaluated by the PAR, defined as the maximum aggregated load (i.e., peak load) to average load used by consumer. The PAR values is reduced to 15\% and 5.8\%  by GA and HSA, respectively, compared to the scenario i.e., "Without HEMS". Moreover, in  Fig.~\ref{fig:3} (left side) , we demonstrate the PAR for scenarios ;multiple users (10 and 50) with the time resolution of 60 and 30 minutes. Out of all maximum reduction of PAR is 17.41 \%, which can be seen for 50 homes with time interval 60 minute. Thus, It shows that the energy consumption is shifted from peak to off- peak time period, proving that GA-HEMS and HSA-HEMS can manage energy consumption adeptly .
%
%

Fig.~\ref{fig:3} (right side) presents the comparison of electricity. It can be observed that the deployment of HSA-HEMS and GA-HEMS reduces the cost in contrast to the case "Without HEMS". Since both algorithms attempt to reduce the cost of the electricity, however, among all (a), (b), (c), and (d) the maximum cost is reduced 1.83\% by HSA-HEMS in case of 50 users and 60 minutes time slots.  It is also seen that cost of energy is maximum (13285 cents for 50 users and 30 minutes time slots) for one complete day "Without HEMS", because most of the energy is used either in peak time or mid peak, while on the contrary, HSA-HEMS and GA-HEMS reduce the cost  (1.64\%) and (1.34\%), respectively, in response to pricing tariff. The statistical analysis of the designed scenarios is provided in Table \ref{tab-anali}.

%
%


\section{Conclusion}
In this paper we designed heuristic model combing with DR strategies for optimizing energy consumption in residential sector. Considering HEMS environment, we modeled our optimization problem  using various types of household appliances, electricity pricing tariffs, and energy demand.  The results show that designed algorithms; GA and HSA can effectively optimize energy consumption,  reduce electricity cost by 0.9 \%, 3.98 \%, and PAR by 15 \%, 5.8 \%, respectively. Also with the different number of users and timescales, the relative performance of both algorithms is effective and minimized the cost and PAR accordingly. As a result, efficient management of resources, peak shaving (power grid), and improve energy usage rate of power grid can be achieved. Simulation results illustrate that both heuristics illustrated the potential of those heuristics in terms of electricity cost and PAR. Besides, it is also shown that DR-based strategies encourage the consumer to manage their energy consumption by shifting the peak hours into off peaks.

In future works, we expect to include distributed energy resources and also incorporate pollution emitted at time of electricity generation, then it would become multi-objective problem (cost and pollution minimization). We also plan to analyze the impact of cyber-attacks in the price signals used by the HEMS.

\section*{Acknowledgements}
This paper is supported by Academy of Finland via: (a) ee-IoT project n.319009, (b) FIREMAN consortium CHIST-ERA/n.326270, and (c) EnergyNet Research Fellowship n.321265/n.328869.

\bibliographystyle{IEEEtran}
\bibliography{ref}

\begin{thebibliography}{10}
\providecommand{\url}[1]{#1}
\csname url@samestyle\endcsname
\providecommand{\newblock}{\relax}
\providecommand{\bibinfo}[2]{#2}
\providecommand{\BIBentrySTDinterwordspacing}{\spaceskip=0pt\relax}
\providecommand{\BIBentryALTinterwordstretchfactor}{4}
\providecommand{\BIBentryALTinterwordspacing}{\spaceskip=\fontdimen2\font plus
\BIBentryALTinterwordstretchfactor\fontdimen3\font minus
  \fontdimen4\font\relax}
\providecommand{\BIBforeignlanguage}[2]{{%
\expandafter\ifx\csname l@#1\endcsname\relax
\typeout{** WARNING: IEEEtran.bst: No hyphenation pattern has been}%
\typeout{** loaded for the language `#1'. Using the pattern for}%
\typeout{** the default language instead.}%
\else
\language=\csname l@#1\endcsname
\fi
#2}}
\providecommand{\BIBdecl}{\relax}
\BIBdecl

\bibitem{01}
R.~Lu, S.~H. Hong, and M.~Yu, ``Demand response for home energy management
  using reinforcement learning and artificial neural network,'' \emph{IEEE
  Transactions on Smart Grid}, 2019.

\bibitem{02}
M.~H. Albadi and E.~F. El-Saadany, ``A summary of demand response in
  electricity markets,'' \emph{Electric power systems research}, vol.~78,
  no.~11, pp. 1989--1996, 2008.

\bibitem{03}
J.~S. Vardakas, N.~Zorba, and C.~V. Verikoukis, ``A survey on demand response
  programs in smart grids: Pricing methods and optimization algorithms,''
  \emph{IEEE Communications Surveys \& Tutorials}, vol.~17, no.~1, pp.
  152--178, 2014.

\bibitem{04}
H.~Shareef, M.~S. Ahmed, A.~Mohamed, and E.~Al~Hassan, ``Review on home energy
  management system considering demand responses, smart technologies, and
  intelligent controllers,'' \emph{IEEE Access}, vol.~6, pp. 24\,498--24\,509,
  2018.

\bibitem{05}
F.-L. Meng and X.-J. Zeng, ``A profit maximization approach to demand response
  management with customers behavior learning in smart grid,'' \emph{IEEE
  Transactions on Smart Grid}, vol.~7, no.~3, pp. 1516--1529, 2015.

\bibitem{06}
W.~Fan, N.~Liu, and J.~Zhang, ``Multi-objective optimization model for energy
  mangement of household micro-grids participating in demand response,'' in
  \emph{2015 IEEE Innovative Smart Grid Technologies-Asia (ISGT ASIA)}.\hskip
  1em plus 0.5em minus 0.4em\relax IEEE, 2015, pp. 1--6.

\bibitem{07}
A.~Khalid, N.~Javaid, M.~Guizani, M.~Alhussein, K.~Aurangzeb, and M.~Ilahi,
  ``Towards dynamic coordination among home appliances using multi-objective
  energy optimization for demand side management in smart buildings,''
  \emph{Ieee Access}, vol.~6, pp. 19\,509--19\,529, 2018.

\bibitem{08}
M.~Awais, N.~Javaid, K.~Aurangzeb, S.~Haider, Z.~Khan, and D.~Mahmood,
  ``Towards effective and efficient energy management of single home and a
  smart community exploiting heuristic optimization algorithms with critical
  peak and real-time pricing tariffs in smart grids,'' \emph{Energies},
  vol.~11, no.~11, p. 3125, 2018.

\bibitem{09}
K.~Ma, T.~Yao, J.~Yang, and X.~Guan, ``Residential power scheduling for demand
  response in smart grid,'' \emph{International Journal of Electrical Power \&
  Energy Systems}, vol.~78, pp. 320--325, 2016.

\bibitem{10}
Z.~Zhao \emph{et~al.}, ``An optimal power scheduling method for demand response
  in home energy management system,'' \emph{IEEE Transactions on Smart Grid},
  vol.~4, no.~3, pp. 1391--1400, 2013.

\bibitem{11}
C.~Bharathi, D.~Rekha, and V.~Vijayakumar, ``Genetic algorithm based demand
  side management for smart grid,'' \emph{Wireless Personal Communications},
  vol.~93, no.~2, pp. 481--502, 2017.

\bibitem{12}
D.~Mahmood, N.~Javaid, N.~Alrajeh, Z.~Khan, U.~Qasim, I.~Ahmed, and M.~Ilahi,
  ``Realistic scheduling mechanism for smart homes,'' \emph{Energies}, vol.~9,
  no.~3, p. 202, 2016.

\bibitem{13}
M.~S. Ahmed, A.~Mohamed, T.~Khatib, H.~Shareef, R.~Z. Homod, and J.~A. Ali,
  ``Real time optimal schedule controller for home energy management system
  using new binary backtracking search algorithm,'' \emph{Energy and
  Buildings}, vol. 138, pp. 215--227, 2017.

\bibitem{14}
K.~Wang, H.~Li, S.~Maharjan, Y.~Zhang, and S.~Guo, ``Green energy scheduling
  for demand side management in the smart grid,'' \emph{IEEE Transactions on
  Green Communications and Networking}, vol.~2, no.~2, pp. 596--611, 2018.

\bibitem{15}
H.~Yang, J.~Zhang, J.~Qiu, S.~Zhang, M.~Lai, and Z.~Y. Dong, ``A practical
  pricing approach to smart grid demand response based on load
  classification,'' \emph{IEEE Transactions on Smart Grid}, vol.~9, no.~1, pp.
  179--190, 2016.

\bibitem{19}
C.~{Lin}, D.~{Deng}, W.~{Liu}, and L.~{Chen}, ``Peak load shifting in the
  internet of energy with energy trading among end-users,'' \emph{IEEE Access},
  vol.~5, pp. 1967--1976, 2017.

\bibitem{16}
H.~Hussain, N.~Javaid, S.~Iqbal, Q.~Hasan, K.~Aurangzeb, and M.~Alhussein, ``An
  efficient demand side management system with a new optimized home energy
  management controller in smart grid,'' \emph{Energies}, vol.~11, no.~1, p.
  190, 2018.

\bibitem{20}
I.~Lütkebohle, ``{Smart Meter Upgrade The Customer-Led Transition to
  Time-of-Use},''
  \url{https://www.cru.ie/wp-content/uploads/2018/05/CRU19019-Customer-Led-Transition-to-Time-of-Use.pdf/},
  2019.

\bibitem{21}
L.~Alberini, ``{As energy gets smarter, ‘time of use’ tariffs spread
  globally},''
  \url{https://www.smart-energy.com/industry-sectors/electric-vehicles/as-energy-gets-smarter-time-of-use-tariffs-spread-globally/},
  2019.

\bibitem{22}
``{Time-of-Use (TOU) Pricing and Schedules},''
  \url{https://www.powerstream.ca/customers/rates-support-programs/time-of-use-pricing.html},
  2019.

\bibitem{23}
J.~H. Holland \emph{et~al.}, \emph{Adaptation in natural and artificial
  systems: an introductory analysis with applications to biology, control, and
  artificial intelligence}.\hskip 1em plus 0.5em minus 0.4em\relax MIT press,
  1992.

\bibitem{24}
O.~Abdel-Raouf and M.~A.-B. Metwally, ``A survey of harmony search algorithm,''
  \emph{International Journal of Computer Applications}, vol.~70, no.~28, 2013.

\bibitem{25}
X.-S. Yang, ``Harmony search as a metaheuristic algorithm,'' in
  \emph{Music-inspired harmony search algorithm}.\hskip 1em plus 0.5em minus
  0.4em\relax Springer, 2009, pp. 1--14.

\end{thebibliography}

\end{document}